\documentclass[10pt,conference]{IEEEtran}
\IEEEoverridecommandlockouts

\usepackage{cite}
\usepackage{amsmath,amssymb,amsfonts}
\usepackage{algorithmic}
\usepackage{textcomp}
\usepackage{xcolor}

\usepackage{caption}
\usepackage{graphicx}
\usepackage{float} 
\usepackage{subfigure}

\def\BibTeX{{\rm B\kern-.05em{\sc i\kern-.025em b}\kern-.08em
    T\kern-.1667em\lower.7ex\hbox{E}\kern-.125emX}}

\begin{document}


\title{Passive Eavesdropping Can Significantly Slow Down RIS-Assisted Secret Key Generation}


\author{
\thanks{*Xiaofeng Tao is the corresponding author}
\IEEEauthorblockN{
Ningya Xu,
Guoshun Nan,
Xiaofeng Tao\IEEEauthorrefmark{1}
\IEEEauthorblockA{Beijing University of Posts and Telecommunications}
\IEEEauthorblockA{xuningya2017@bupt.edu.cn, nanguo2021@bupt.edu.cn, taoxf@bupt.edu.cn}
}
}

\maketitle

\begin{abstract}
Reconfigurable Intelligent Surface (RIS) assisted physical layer key generation has shown great potential to secure wireless communications by smartly controlling signals such as phase and amplitude. However, previous studies mainly focus on RIS adjustment under ideal conditions, while the correlation between the eavesdropping channel and the legitimate channel, a more practical setting in the real world, is still largely under-explored for the key generation. To fill this gap, this paper aims to maximize the RIS-assisted physical-layer secret key generation by optimizing the RIS units switching under the eavesdropping channel. Firstly, we theoretically show that passive eavesdropping significantly reduces RIS-assisted secret key generation. Keeping this in mind, we then introduce a mathematical formulation to maximize the key generation rate and provide a step-by-step analysis. Extensive experiments show the effectiveness of our method in benefiting the secret key capacity under the eavesdropping channel. We also observe that the key randomness, and unmatched key rate, two metrics that measure the secret key quality, are also significantly improved, potentially paving the way to RIS-assisted key generation in real-world scenarios.   
\end{abstract}

\begin{IEEEkeywords}
secret key generation, physical layer, reconfigurable intelligent surface
\end{IEEEkeywords}

\section{Introduction}

Conventional encryption mechanisms based on public-private keys can secure wireless networks at the upper layer\cite{sgk}. 
While these encryption methods require secret keys that are available only between legitimate parties, distributing the keys will introduce additional computation and communication costs. Thus, deployment of the above approaches will be challenging, especially on resource-constrained large-scale mobile networks, such as the Internet of Things (IoT) and Machine-to-Machine communications (M2M). The newly emerged physical-layer secret key technology, a promising physical-layer security mechanism, provides lightweight encryption \cite{lgy1} for securing wireless communications. The underlying principle is to extract natural random sources using endogenous security elements\cite{hkz} of wireless networks, such as the time-variability of wireless channels. However, the key generation rate, vital to physical-layer encryption, requires rapidly changing the wireless channel state information (CSI) to maximize the channel entropy. While in the real world, the mobile 
clients may be located in quasi-static environment\cite{lgy2}, such as smart home and environmental monitoring, and pose great challenges for producing secret keys.

To apply physical-layer encryption in various scenarios, we call for a method that can mitigate the impact of quasi-static wireless environments on the key generation. Reconfigurable Intelligent Surface (RIS) has attracted increasing attention as it can smartly control the radio signals between a transmitter and a receiver in a dynamic and goal-oriented way.
Although effective, the complexity of the RIS system is also much lower than traditional relay systems\cite{wq}, as RIS only requires local coverage without any radio frequency (RF) links. 

Prior effects mainly focus on how RIS automatically customizes the wireless transmission environment to maximize the key generation rate\cite{hyn,jzj1,hxy,lt}. However, these methods work under ideal conditions, assuming that the illegal eavesdropper is always located half wavelength away from the legitimate user. Thus it will be hard to introduce impact on the randomness of the wireless channel, while the correlation between the eavesdropping channel and the legitimate channel - a more practical setting in the real world - may lead to low key generation rate with only 1 bit/s\cite{pierrot}. Hong\cite{Hong} proposed a key generation scheme that superimposes the artificial noise orthogonal to the legal channel, aiming to interfere with eavesdropped signals. But such an act will inevitably introduce additional transmission power, which is not suitable for RIS-assisted scenarios with power-consuming characteristics\cite{zuo}\cite{zhang}, the corresponding solutions still need further exploring.

In order to fill in the gap of RIS control method optimization in physical layer key generation under related eavesdropping channels, this paper proposes a RIS dynamic control strategy to maximize secret key capacity. Specifically, there are two analysis steps involved in our method: 1) under the channel state information(CSI) knowable hypothesis, all sub-channels CSI can be obtained through channel estimation and used to analyze the influence of eavesdropping on key capacity performance when the eavesdropper is close to the legitimate sender and RIS respectively; 2) real-time CSI is used to control the switching state of each component of the RIS to obtain the key rate gain. 

The main contributions of this paper are as follows:\\
$\bullet$ We present a novel RIS configuration method which optimizes the exact switched-on location of RIS by using real-time CSI under a relevant eavesdropping channel, to enhance the key capacity even when RIS resources are limited.\\
$\bullet$ We introduce a novel process of calculating secret key capacity under eavesdropping, which can adjust the expression with the consideration of analyzing different eavesdropping cases.\\
$\bullet$ We conduct extensive experiments to verify the effectiveness of our method, and the results show outstanding performance compared with the random switching method.

\section{Related Work}

\subsection{RIS assisted wireless communication scenarios}

RIS-assisted wireless communication systems \cite{wq} have received extensive academic attention, including hardware development and performance optimization\cite{lhc}. Our paper regulates the switching characteristics of RIS, rather than the phase or amplitude of RIS components used in most studies. Compared with the simple control RIS components\cite{xl}\cite{jt}, the uncertainty brought by the random switched units of RIS is more suitable for the presence of eavesdropping. Moreover, RIS units switching method is more practical to be used in the actual situation that RIS resources are limited.

\subsection{RIS assisted secret key generation}

Hu Y.\cite{hyn} and Ji\cite{jzj1} designed a heuristic algorithm framework to manipulate the switching of RIS phase shift matrix. Hu X.\cite{hxy} considered optimizing the phase and amplitude of RIS through SDR/SCA algorithms. In \cite{lt}, it was found that RIS component switching time could be searched to improve key generation rate. Qin\cite{zr}\cite{qq} and Nan\cite{ngs} proposed using deep learning-based methods such as semantic communication to further strengthen the robustness of RIS-assisted secret key system. But all the above studies assume that the illegal eavesdropper locate half wavelength away from the legitimate user, and our paper considers the influence of eavesdropping channel on the sum secret key capacity.

\section{System Model}

\subsection{Secret key generation procedure}

\begin{figure}[htbp]
\centerline{\includegraphics[width=0.9\linewidth]{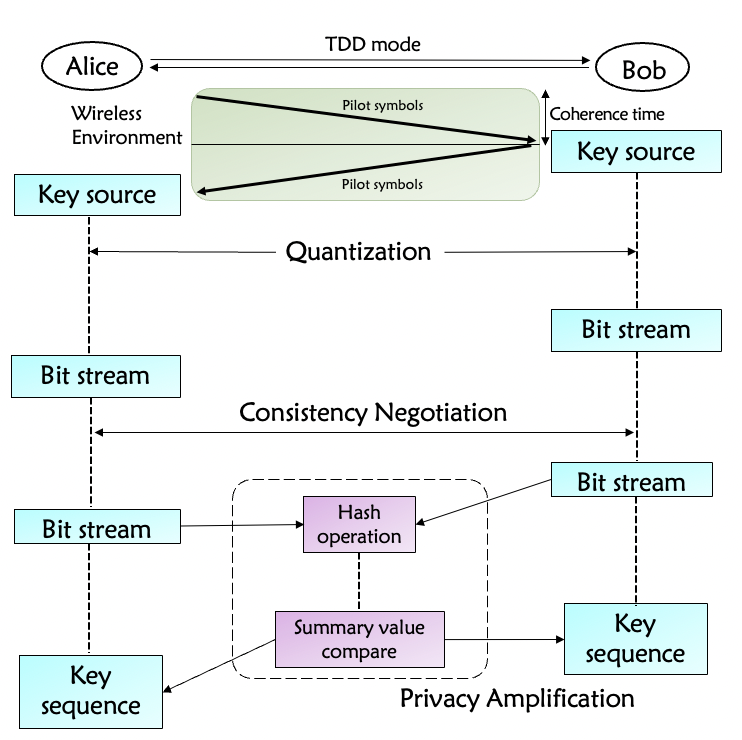}}
\caption{Procedure of secret key generation from wireless channels.}
\label{2-1}
\end{figure}

The typical process of obtaining shared secret key in current research is shown in Fig.~\ref{2-1}, which mainly includes the following four steps\cite{cw}:

\paragraph{Channel Measurement} Alice and Bob send pilot frequency to the peer end, and generate random key sources by probing some characteristics of the channel, e.g. received signal strength(RSS), CSI, channel phase response and channel multi-path delay.

\paragraph{Quantization} The communication parties convert the values obtained through channel measurement into a bit sequence of 0,1.

\paragraph{Consistency Negotiation} Use an information reconciliation protocol to discard or correct the difference between the key bit stream and reduce the inconsistency rate.

\paragraph{Privacy Amplification} Discard some inconsistent bits or perform some bit conversion to strengthen the key, obscure the local information that the eavesdropper may obtain in the Consistency Negotiation step.

As our paper researches on the performance of our RIS configuration approach brought to the secret key capacity, we focus mostly on the \textit{Channel Measurement} step and adopt CSI as key source.

\subsection{Our RIS assisted system model}

\begin{figure}[htbp]
\centerline{\includegraphics[width=0.8\linewidth]{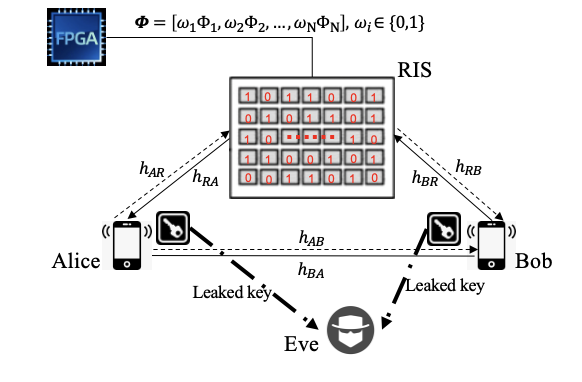}}
\caption{Secret key generation based on RIS assistance.}
\label{2-2}
\end{figure}

The RIS-assisted wireless key generation system model is shown in Fig.~\ref{2-2}. Alice and Bob are the legitimate communication parties who can extract random keys from the wireless channel information they observe, an illegal eavesdropper Eve who tries to eavesdrop on the key generated by the legal two parties, but cannot actively interfere with them. Alice, Bob and Eve are all single-antenna devices. The system adopt a time-division duplex system, to ensure that the up-down channels meet the reciprocity in a coherent time. 

A RIS equipped with $N$ reflection units is located between Alice and Bob to enhance the channel randomness, which can be programmed through the wired link of the controller. Assume that all reflection units of RIS are independent, and the status value $\omega$ can be set to ``on" $\omega=1$ or ``off" $\omega=0$ by the controller \cite{xl}. By controlling the status value of RIS units, we change the phase shift matrix $\mathbf\varPhi=[\omega_1\varPhi_1,\omega_2\varPhi_2,\omega_3\varPhi_3,...,\omega_N\varPhi_N],\, \omega_i\in\{0, 1\}$ in real time, where $\varPhi_i$ denotes the random phase shift corresponding to each RIS unit. 

Set that the wireless channel between each nodes are $h_j\sim CN(0,\sigma^2_{h_j}), j\in\{AB,BA,AE,BE\}$,  which are all modeled as quasi-static block fading channel satisfying the complex Gaussian distribution of zero mean. What's more, take upstream channel as example, the channel coefficients between each nodes and RIS can be expressed as $h_{AR}\in \mathbb C^{1\times N}$, $h_{RB}\in \mathbb C^{N\times 1}$, $h_{RE}\in \mathbb C^{N\times 1}$, and vice versa. Each element in the channel coefficients matrix follows the Gaussian distribution.

The pilot sequence $s$ sent by Alice and Bob is multiplied by the total phase shift matrix $\mathbf\varPhi$, then consider channel estimation method by using least square(LS) method to the received signal strength, the CSI at Alice and Bob could be expressed as $H_A$ \eqref{1a} and $H_B$ \eqref{1b}, where $n_A$ and $n_B$ denote the Gaussian White Noise.

\begin{subequations}
\begin{small}
\begin{equation}
\begin{aligned}
    H_A= h_{BA}+\sum_{i=1}^N h_{BR}^i \omega_i\varPhi_i  h_{RA}^i +n_A
\end{aligned}
\label{1a}
\end{equation}
\begin{equation}
\begin{aligned}
    H_B= h_{AB}+\sum_{i=1}^N h_{AR}^i \omega_i\varPhi_i  h_{RB}^i +n_B
\end{aligned}
\label{1b}
\end{equation}
\end{small}
\end{subequations}

Based on the above model, we will discuss the eavesdropping strategy when Eve is close to one legitimate user and RIS respectively. By analyzing the influence of Eve's position on secret key capacity, the switching strategy of RIS in the case of extreme eavesdropping situation will be found, so as to maximize the upper bound of key capacity and enhance the secrecy performance.

\section{The proposed Approach}

\subsection{Different eavesdropping scenarios analysis}

Based on the definition of physical layer secret key generation \cite{maurer}, the secret key capacity $C_{SK}$ can be expressed as conditional mutual information of CSI in equation \eqref{2}, where the CSI observed at Alice and Bob can be given in Section \uppercase\expandafter{\romannumeral3}.B. More specifically, the expression for $H_E$ includes both the channel observations between Alice-Eve $H_{AE}$ \eqref{3a}  and Bob-Eve $H_{BE}$ \eqref{3b} , which varies depending on the location of Eve. 
\begin{equation}
\begin{aligned}
    C_{SK}&=I(H_A;H_B|H_E)\\
    &=I(H_A;H_B|H_{AE},H_{BE})
\end{aligned}
\label{2}
\end{equation}

\begin{subequations}
\begin{small}
\begin{equation}
\begin{aligned}
    H_{AE}= h_{AE}+\sum_{i=1}^N h_{AR}^i \omega_i\varPhi_i  h_{RE}^i +n_{AE}
\end{aligned}
\label{3a}
\end{equation}
\begin{equation}
\begin{aligned}
    H_{BE}= h_{BE}+\sum_{i=1}^N h_{BR}^i \omega_i\varPhi_i  h_{RE}^i +n_{BE}
\end{aligned}
\label{3b}
\end{equation}
\end{small}
\end{subequations}

The best eavesdropping attack strategy for Eve is to approach one legitimate node or RIS itself to eavesdrop, so we will talk about two main eavesdropping cases: Eve close to one legitimate node and Eve close to RIS. In order to better express the RIS auxiliary key generation performance under eavesdropping, it is assumed that the correlation factor between the eavesdropping channel and the legitimate channel is denote as $\rho$, which is positively correlated with the distance between Eve and the observed legitimate node\cite{jt}. To be more precisely, Let the correlation coefficient $\rho = J_0 (2 \pi l / \lambda) \in [0,1]$, where $J_0$ is the zero-order Bessel function, $l$ is the distance between Eve and the observed legitimate node, and $\lambda$ is the wavelength of the carrier signal.\\

\noindent $\bullet$ Eve is close to one legitimate node

\begin{figure}[htbp]
\centerline{\includegraphics[width=0.8\linewidth]{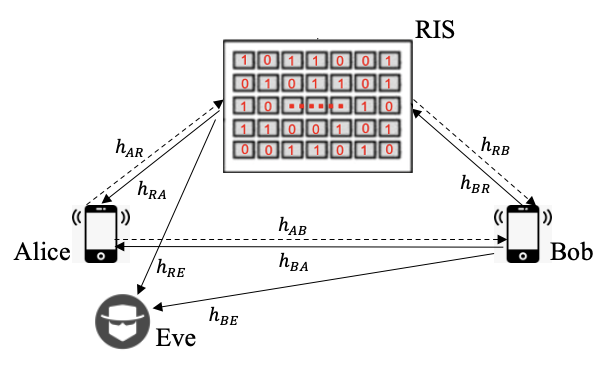}}
\caption{Eve is close to one legitimate node}
\label{3-1}
\end{figure}

Because of the channel symmetry property, the analysis of Alice and Bob are the same, so we choose to analyze Alice. We assume that Eve can only eavesdrop on the node that it is close to. 
\begin{equation}
\begin{aligned}
    h_{BE}=\rho h_{BA}+\sqrt{1-\rho^2}n ,\ h_{RE}=\rho h_{RA}+\sqrt{1-\rho^2}n
\end{aligned}
\label{4}
\end{equation}

When $\rho \to 1$, it's obvious to see that $H_{AE}$ is independent with the legitimate channels because the correlation between $h_{RE}$ and $h_{RA}$ becomes stronger with the movement of Eve. Without the uncertainty of Gaussian background noise, Eve can get almost all the information about $H_A$ from the $H_{BE}$ in \eqref{5}.
\begin{equation}
\begin{small}
\begin{aligned}
    H_{BE}&=h_{BE}+\sum_{i=1}^N h_{BR}^i \omega_i\varPhi_i  h_{RE}^i +n_E\\
    &\approx h_{BA}+\sum_{i=1}^N h_{BR}^i \omega_i\varPhi_i  h_{RA}^i +n_A=H_A
\end{aligned}
\end{small}
\end{equation}

So the secret key capacity could be simplified as equation \eqref{6}, which matches the setting that Eve can only eavesdrop on the node that it is close to.
\begin{equation}
    C_{SK}=I(H_A;H_B|H_{BE})
\label{6}
\end{equation}

\noindent $\bullet$ Eve is close to RIS

\begin{figure}[htbp]
\centerline{\includegraphics[width=0.8\linewidth]{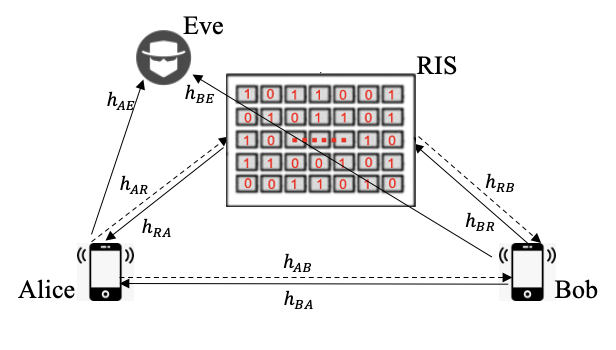}}
\caption{Eve close to RIS}
\label{3-2}
\end{figure}

When Eve is closer to RIS, it is far more than 1/2 wavelength away from both Alice and Bob. The correlation between the legitimate node-Eve channel and the legitimate node-RIS channel becomes stronger.
\begin{equation}
\begin{small}
\begin{aligned}
    h_{AE}=\rho h_{AR}+\sqrt{1-\rho^2}n  ,\ h_{BE}=\rho h_{BR}+\sqrt{1-\rho^2}n
\end{aligned}
\end{small}
\end{equation}

As $\rho \to 1$, it can be inferred from the expression of eavesdropping channels that $H_{AE} $ and $H_{BE} $ remain approximately independent of $H_A$ and $H_B$, which makes it difficult for Eve to get key bit information from its own channel estimation matrix. So the secret key capacity result is approximately the same as the case of an independent eavesdropping channel, which is shown in equation \eqref{8}.
\begin{equation}
    C_{SK}=I(H_A;H_B)
\label{8}
\end{equation}

\subsection{RIS units switching method}

From the secret key capacity results we just analyzed, to deal with the more complicated eavesdropping case, we adopt the analysis based on the case of Eve close to one legitimate node. After expanding based on the conditional mutual information theorem, we see that the expression of secret key capacity $C_{SK}$ can be written as a form of determinant of covariance matrix in equation \eqref{9}: 
\begin{equation}
 \begin{aligned}
    C_{SK}&=I(H_A;H_B|H_{BE})\\
    &=I(H_A;H_{BE})-I(H_A;H_B, H_{BE})\\
    &=\log_2 \frac{det(R(H_A,H_{BE}))\times det(R(H_B,H_{BE}))}{det(R(H_{BE})\times det(R(H_A,H_B,H_{BE}))}
\end{aligned}
\label{9}
\end{equation}

Whereas the covariance matrix of multiple matrices $A_1 \cdots A_n$ can be expressed by the cross entropy of vector $a_1 \cdots a_n$ in equation \eqref{10}, which are corresponding vectors after the matrices vectorization.
\begin{equation}
R(A_1, ..., A_n)=E \begin{bmatrix}
a_1a^*_1 &\cdots &a_na^*_1\\ \vdots &\ddots &\vdots\\ a_1a^*_n &\cdots &a_na^*_n 
\end{bmatrix}
\label{10}
\end{equation}

As we already exclude the influence factor of Eve with the node that is not close to it, there are totally 3 channels $H_A$, $H_B$, $H_{BE}$ into account. All 3 channels are cascaded channels of RIS-assisted channel and direct channel with the background noise. Take $H_{BE}$ as an example, the covariance matrix and its determinant can be expressed by the combination of sub-channels in \eqref{11}. 
\begin{equation}
\begin{small}
 \begin{aligned}
    R(H_{BE})=\rho R(H_{AB})+\rho R(\sum_{i=1}^N h_{RA}^i \omega_i h_{BR}^i )+\sigma^2_n I
\end{aligned}
\end{small}
\label{11}
\end{equation}

Suppose that the noise power $\sigma^2_n =1$ for all channels. Since the real and imaginary parts of all elements matrices are subject to independent and identically distributed Gaussian variables, all 4 terms can be written as a form of the power of different sub-channels in \eqref{12a}, \eqref{12b}, \eqref{12c}:  

\begin{subequations}
\begin{footnotesize}
\begin{equation}
 \begin{aligned}
        detR(H_{BE})= \rho \left[\sigma^2_{AB}+\sum_{i=1}^N \frac{\omega^2_i(\sigma^2_{RA}\sigma^2_{RB})}{\sigma^2_{RA}+\sigma^2_{RB}}\right]+1
\end{aligned}
\label{12a}
\end{equation}

\begin{equation}
 \begin{aligned}
    detR(H_A, H_B, H_{BE})=(\rho^2+2)\left[\sigma^2_{AB}+\sum_{i=1}^N \frac{\omega^2_i(\sigma^2_{RA}\sigma^2_{RB})}{\sigma^2_{RA}+\sigma^2_{RB}}\right]+1
 \end{aligned}
 \label{12b}
\end{equation}

\begin{equation}
 \begin{aligned}
    detR(H_A, H_{BE})&=detR(H_B, H_{BE})\\
    &=(\rho^2+1)\left[\sigma^2_{AB}+\sum_{i=1}^N \frac{\omega^2_i(\sigma^2_{RA}\sigma^2_{RB})}{\sigma^2_{RA}+\sigma^2_{RB}}\right]+1
 \end{aligned}
 \label{12c}
\end{equation}

\end{footnotesize}
\end{subequations}

For simplicity, we assume that $\sigma^2_{AB}+\sum_{i=1}^N \frac{\omega^2_i(\sigma^2_{RA}\sigma^2_{RB})}{\sigma^2_{RA}+\sigma^2_{RB}}$ to be $x$, so the secret capacity final expression is as shown in equation \eqref{13}. It is verified that when $\rho=0$, the key rate expression is the same as the expression without considering the eavesdropping channel (that is, the eavesdropping channel is independent of the legal channel), which proves the correctness of the result.
\begin{equation}
\begin{small}
 \begin{aligned}
    C_{SK}=\log_2\frac{(\rho^4+2\rho^2+1)x^2+(2\rho^2+4\rho+2)x+1}{(\rho^3+2\rho)x^2+(\rho^2+\rho+2)x+1}
\end{aligned}
\end{small}
\label{13}
\end{equation}

Since our goal is to find an optimal RIS configuration method to obtain the maximum secret key capacity, the final optimization expression is shown below. Based on this optimization function, the best RIS unit placement location $\omega$ corresponding to the maximum variance item of RIS cascade channel $\frac{(\sigma^2_{RA}\sigma^2_{RB})}{\sigma^2_{RA}+\sigma^2_{RB}}$ is found and placed in the open state, to maximize the key capacity $C_{SK}$ when RIS resources are limited under $M$.
\begin{equation*}
\begin{split}
&\max_{\omega_i} \,\, C_{sk}\\
&s.t.\quad  \left\{\begin{array}{lc}
\sum_{i=1}^{N} \omega^2_i \le M\\
M \le N\\
0<\rho<1\\
\end{array}\right.
\end{split}
\end{equation*}

\section{Simulation Results}

\begin{figure*}
	\centering
	\subfigure[The influence of correlation factor $\rho$ on secret key capacity $C_{SK}$]{
		\includegraphics[height=5cm, width=5.7cm]{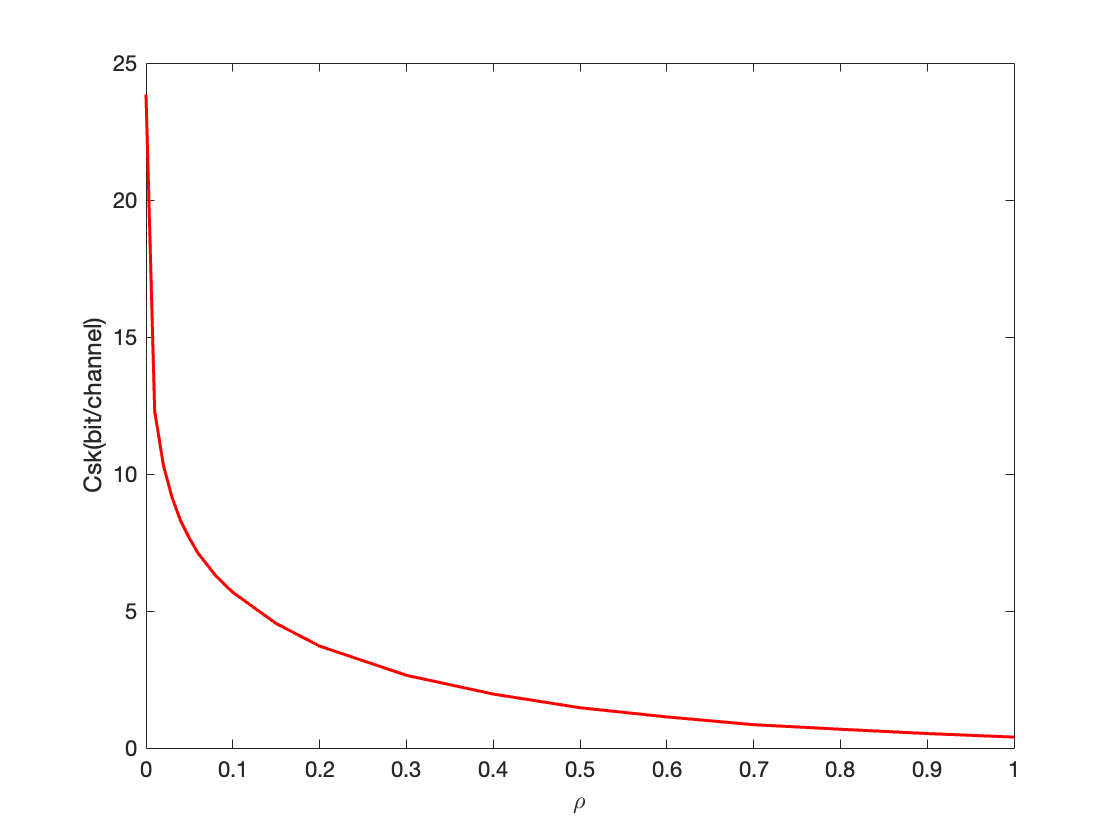}
                \label{5-1}
	}
	\subfigure[$C_{SK}$ comparison on different RIS configuration mode]{
		\includegraphics[height=5cm, width=5.7cm]{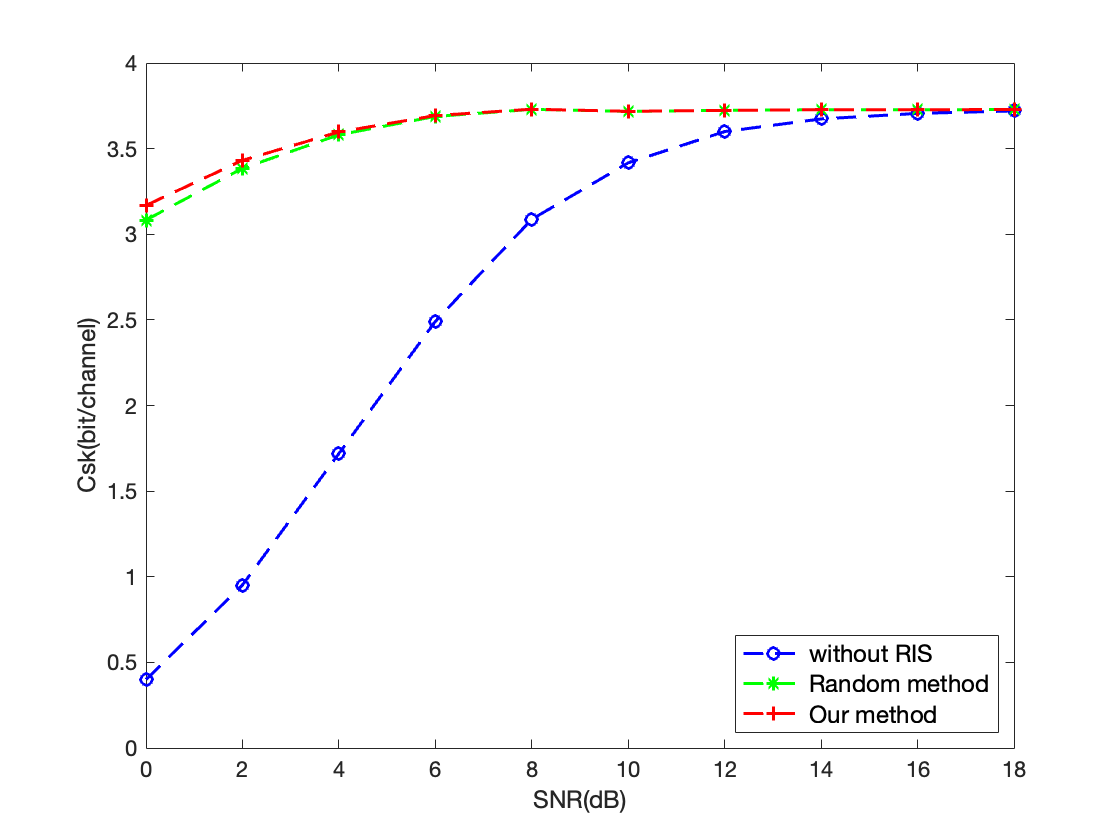}
                \label{5-2}
	}
	\subfigure[$C_{SK}$ comparison under three correlation factors $\rho$]{
		\includegraphics[height=5cm, width=5.7cm]{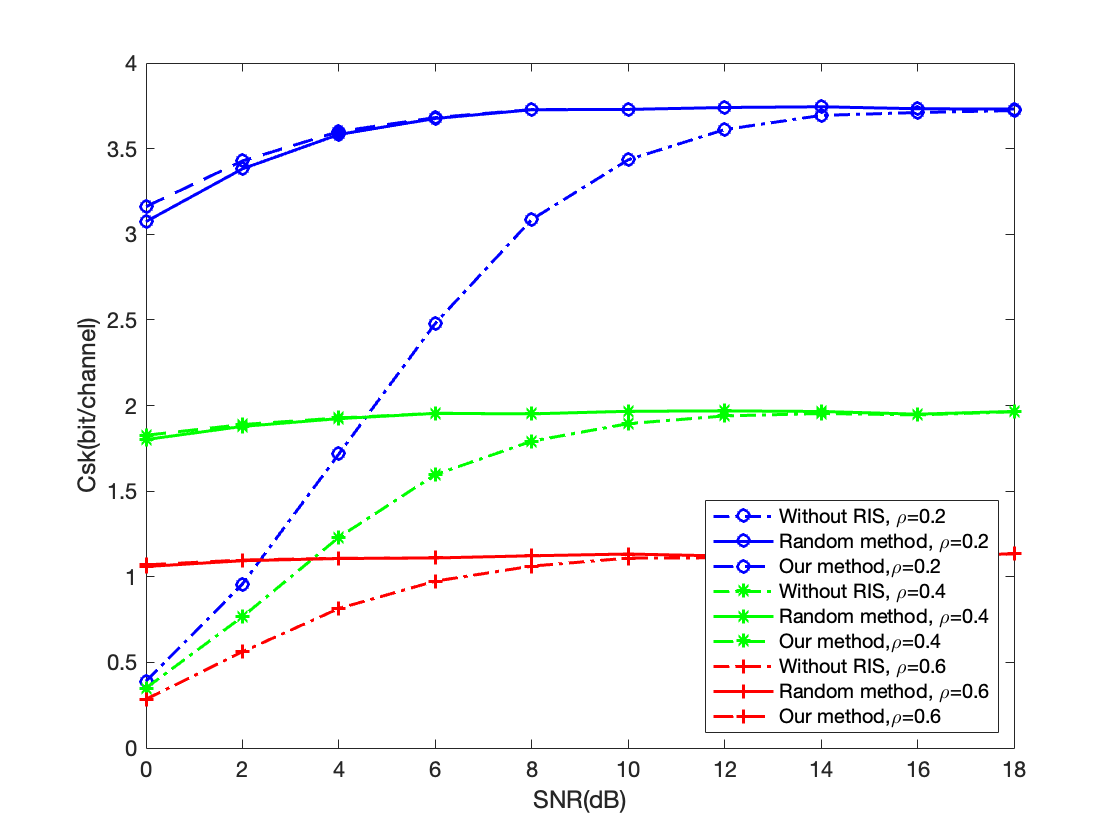}
                \label{5-3}
	}
	\quad
	\subfigure[$C_{SK}$ comparison under three RIS units turn-on rate $r$]{
		\includegraphics[height=5cm, width=5.7cm]{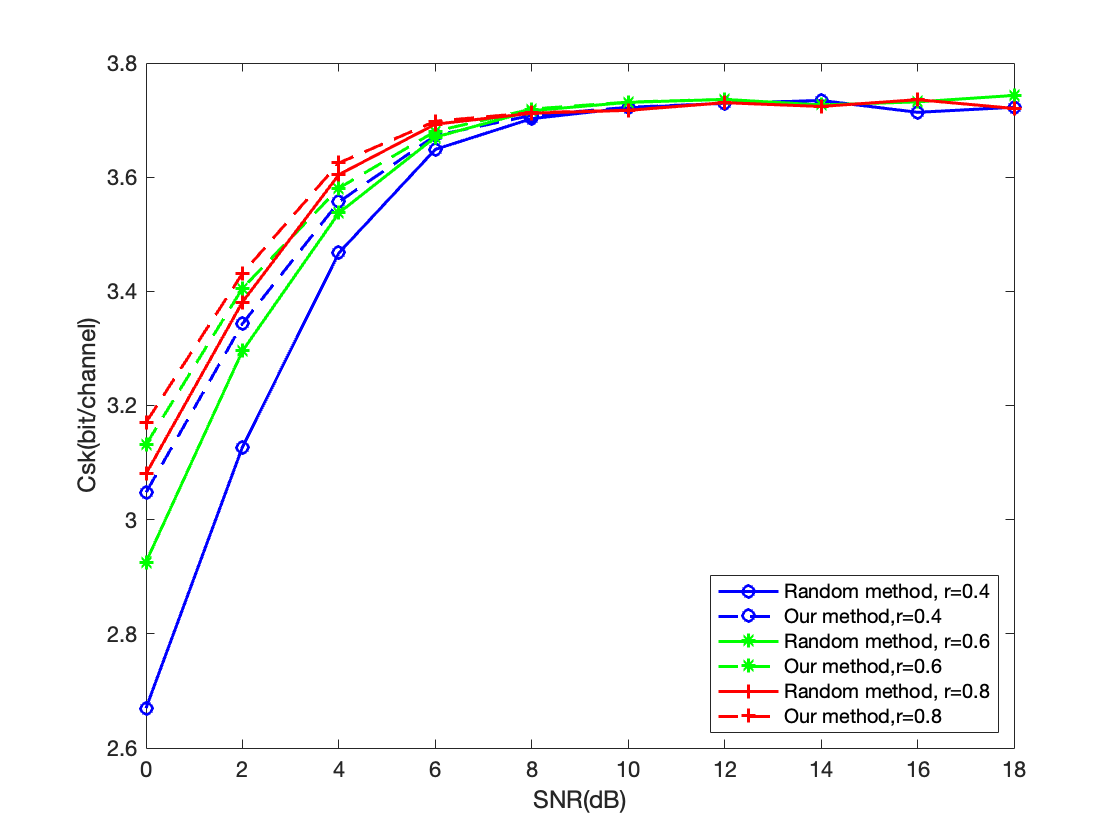}
                \label{5-4}
	}
	\subfigure[$C_{SK}$ comparison under three RIS units number $N$]{
		\includegraphics[height=5cm, width=5.7cm]{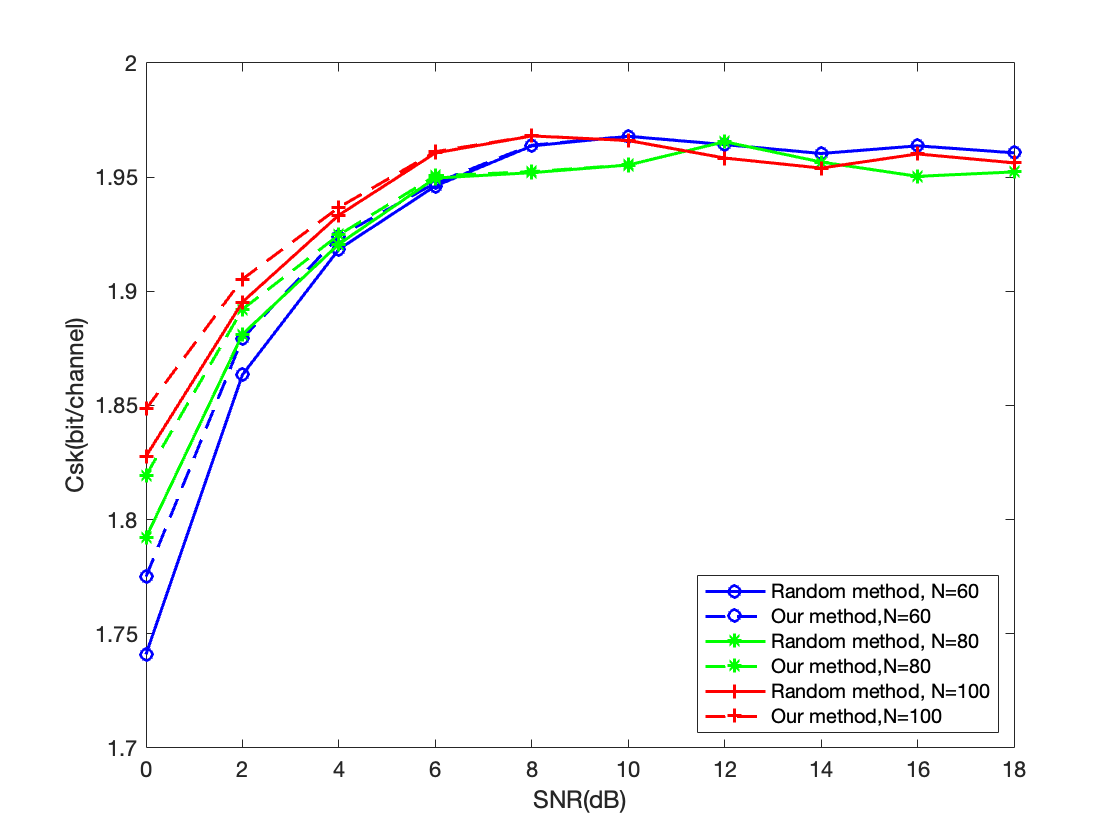}
                \label{5-5}
	}
	\subfigure[unmatched key rate comparison on different RIS configuration mode]{
		\includegraphics[height=5cm, width=5.7cm]{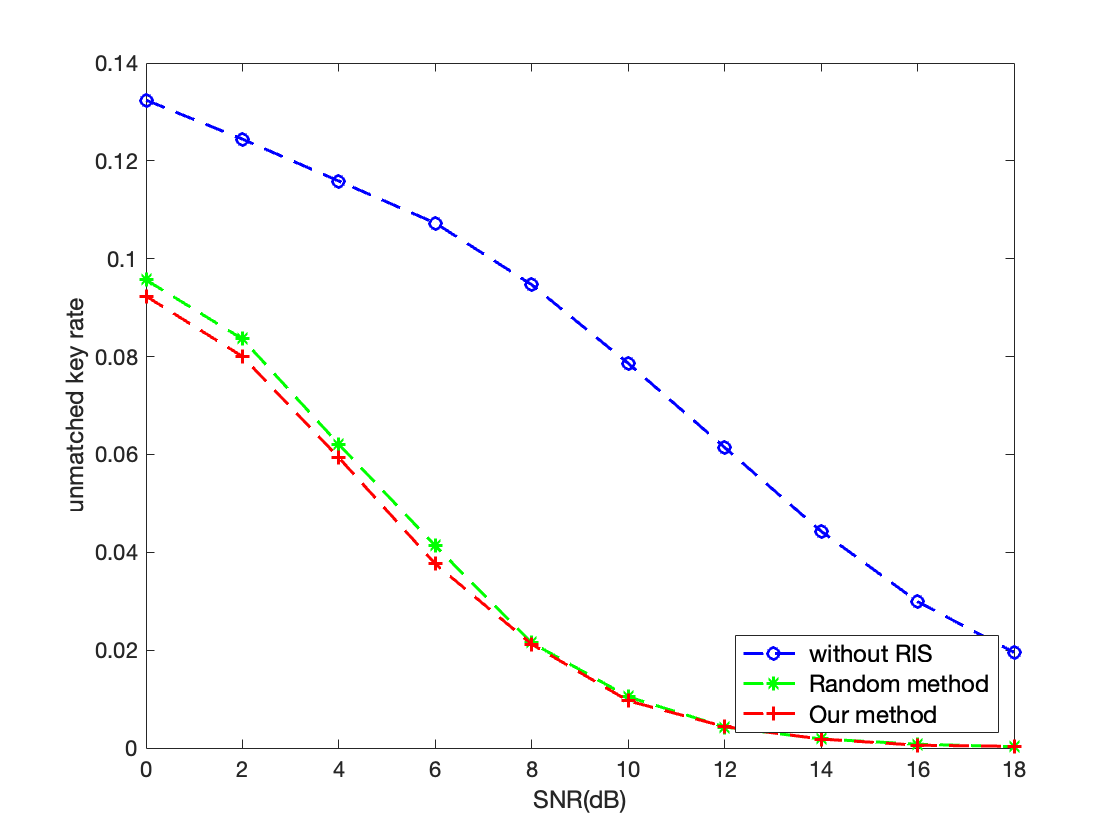}
                \label{5-6}
	}
	\centering
	\caption{Simulation results on our approach. (a) shows how the correlation between eavesdropping and legitimate channels influence the secret key capacity when our approach is adopted. (b) shows the $C_{SK}$ comparison between our approach and two baselines(RIS units random regulation and without RIS), (c),(d),(e) is the expansion based on (b) when different configuration parameters are adopted. (f) shows the unmatched key rate comparison between our approach and two baselines same as in (b).}
	\label{5}
\end{figure*}

In order to verify the performance of the above approach and the correctness of the theoretical analysis, this section conducts simulation experiments based on MATLAB R2021a. Under every 2$dB$ signal-to-noise ratio between 0 to 18$dB$, 100,000 times Monte Carlo experiment are adopted to randomly generate a group of channel matrix values and noise values, and the secret key capacity $C_{SK}$ values are calculated by the result we shown in \eqref{13}.

\subsection{The influence of correlation factor $\rho$}

Fig. \ref{5-1} shows the declining trend of secret key capacity with different correlation factors, the signal-to-noise ratio is set at 18 dB. It can be seen that when $\rho \le 0.1$, the downward trend gradually increases, and the 27 bit/channel without considering the relevance of the eavesdropping channel rapidly drops to 5 bit/channel; while $\rho>0.1$, the downward trend gradually slow down, and approach the lower boundary value of 1.95 bit/channel after $\rho>0.7$. It shows that the correlation between the eavesdropping channel and legitimate channel has a significant influence on the key capacity.

\subsection{The improvement brought to secret key capacity $C_{SK}$}

In order to verify the applicability of our proposed approach, we select the two situations as baselines: without RIS-assisted channel and RIS units random regulation. Through the comparison of $C_{SK}$ of the three RIS unit configured channels participating in key generation, it can be seen from Fig. \ref{5-2} that, under the condition of low signal-to-noise ratio ($SNR<14dB$), the introduction of RIS greatly improves the channel dynamics, thus increasing $C_{SK}$. Moreover, in an environment with large noise pollution ($SNR<6dB$), our method effectively improves the upper bound of $C_{SK}$ compared with RIS units random regulation.

On this basis, we conduct a series of more elaborate tests. First, we test the influence of our method and two baselines on $C_{SK}$ under different correlation factors $\rho$. It can be seen from Fig. \ref{5-3} that the upper bound values of $C_{SK}$ under different correlation factors are different, which is consistent with the verification results in Fig. \ref{5-1}. For each value of $\rho$, our method can obtain the highest $C_{SK}$, and the smaller the $\rho$, the more obvious the effect.

When the number of RIS units is 80, we carry out the test by changing the proportion of RIS units that can be opened with results in Fig. \ref{5-4}. We notice that the application of our approach could significantly improve the $C_{SK}$ compared to RIS units random regulation at the same RIS units opening ratio $r$. According to the simulation data, when $r=0.6$ the $C_{SK}$ obtained by our method exceeds the $C_{SK}$ obtained when $r=0.8$ using the random method. In addition, the lower the opening ratio is, the stronger the improvement effect will be. When $r$ is only 0.4, the $C_{SK}$ applied by our method is close to the $C_{SK}$ with $r=0.8$ under the random method.

Moreover, we also pay attention to the variation of RIS properties brought to $C_{SK}$. Therefore, we initially set the turn-on rate to 0.8, and change the number of RIS units by 60, 80 and 100 respectively to observe the effect. From Fig. \ref{5-5}, when the signal-to-noise ratio is high, the number of RIS units $N$ has a tiny little effect on $C_{SK}$. However, in the case of low signal-to-noise ratio, the larger the $N$, the larger the $C_{SK}$, that is the more keys can be generated in the same period. In addition, with the same number of $N$, $C_{SK}$ can be increased appropriately by applying the RIS configuration method proposed by us rather than the random method.

\subsection{The improvement brought to key consistency rate}

In addition to a significant increase in $C_{SK}$, our RIS regulation approach can also improve key consistency. In the subsequent quantization process, we adopt 2-bit quantization using gray code \cite{hhj} which can effectively limit the inconsistency rate of the quantized sequence because there is only one bit difference between adjacent code words. The unmatched key rate is calculated by dividing the number of inconsistent bits on both sides of the total key sequence. It can be found from Fig. \ref{5-6} that our method significantly reduces the unmatched key rate, which could even be down to zero when the channel condition is well. Our method improves the performance under the condition of low SNR compared with RIS random regulation, which reduces the unmatched key rate by about 3$\%$.

\subsection{Randomness Verification}

In this paper, NIST randomness test is used to evaluate the randomness of keys \cite{bas}. Six randomness test methods are selected to calculate the pass rate of keys generated under different RIS control modes, and 2-bit gray code quantization is also adopted. The results are shown in Table 1. 
\begin{table}[htbp]
\caption{NIST Randomness Test Results}
\begin{center}
\begin{tabular}{|c|c|c|c|}
\hline
\textbf{Statistical}&\multicolumn{3}{|c|}{\textbf{p-value}} \\
\cline{2-4} 
\textbf{Test} & \textbf{\textit{Without RIS}}& \textbf{\textit{Random Method}}& \textbf{\textit{Our Method}} \\
\hline
Frequency$^{\mathrm{a}}$&0.066882 &0.213309 &\textbf{0.350485} \\
\hline
BlockFrequency&0.213309 &0.122325 &0.437274 \\
\hline
Runs&0.637119 &0.964295 &0.834308  \\
\hline
LongestRuns&0.213309 &0.350485 &0.637119 \\
\hline
Serial&0.017912 &0.637119 &0.964295 \\
\hline
LinearComplexity&0.090936 &0.534146 &0.122325 \\
\hline
\multicolumn{4}{l}{$^{\mathrm{a}}$The basis of all following tests.}
\end{tabular}
\label{tab1}
\end{center}
\end{table}

When p-value is higher than 0.01, the randomness test is successfully passed. NIST test results in ``Frequency" show that the randomness of key generation in Without RIS mode is comparatively low because RIS-assisted fast change channel is not introduced, while the randomness in RIS random method and our method is much higher. In addition, compared with RIS units random configuration method, our method has pretty high randomness, which could make the key bit stream less likely to be intercepted by the eavesdropper.

\section{Conclusion and Future Work}

We derived a RIS configuration method to improve the secret key capacity of a RIS-assisted single-antenna system in the presence of an eavesdropper. More specifically, our method used the real-time CSI to control the specific RIS units to open under RIS resources-limited situation, rather than random configuration. Credible numerical results showed the effectiveness of our method. Our proposed method can be further extended to more complex scenarios such as multi-antenna and multi-eavesdropper.

\section{Acknowledgments}
This work was supported by the National Key R\&D Program of China (No. 2022YFB2902200).

\end{document}